\documentclass[preprint]{ptephy}

\preprintnumber{KYUSHU-HET-160}

\usepackage{amssymb}
\usepackage{amsthm}
\usepackage{amsmath}
\usepackage{booktabs}
\usepackage{simplewick}

\numberwithin{equation}{section}

\usepackage{url}

\begin{document}

\title{Upper bound on the mass anomalous dimension in many-flavor gauge
theories: a conformal bootstrap approach}

\author{%
\name{\fname{Hisashi} \surname{Iha}}{1},
\name{\fname{Hiroki} \surname{Makino}}{1}, and
\name{\fname{Hiroshi} \surname{Suzuki}}{1,\ast}
}

\address{%
\affil{1}{Department of Physics, Kyushu University, 744 Motooka, Nishi-ku,
Fukuoka, 819-0395, Japan}
\email{hsuzuki@phys.kyushu-u.ac.jp}
}

\begin{abstract}%
We study four-dimensional conformal field theories with an $SU(N)$ global
symmetry by employing the numerical conformal bootstrap. We consider the
crossing relation associated with a four-point function of a spin~$0$
operator~$\phi_i^{\Bar{k}}$ which belongs to the adjoint representation
of $SU(N)$. For~$N=12$ for example, we found that the theory contains a
spin~$0$ $SU(12)$-breaking relevant operator when the scaling dimension
of~$\phi_i^{\Bar{k}}$, $\Delta_{\phi_i^{\Bar{k}}}$, is smaller than~$1.71$.
Considering the lattice simulation of many-flavor quantum chromodynamics with
$12$~flavors on the basis of the staggered fermion, the above $SU(12)$-breaking
relevant operator, if it exists, would be induced by the flavor-breaking effect
of the staggered fermion and prevent an approach to an infrared fixed point.
Actual lattice simulations do not show such signs. Thus, assuming the absence
of the above $SU(12)$-breaking relevant operator, we have an upper bound on the
mass anomalous dimension at the fixed point~$\gamma_m^*\leq1.29$ from the
relation~$\gamma_m^*=3-\Delta_{\phi_i^{\Bar{k}}}$. Our upper bound is not so
strong practically but it is strict within the numerical accuracy. We also find
a kink-like behavior in the boundary curve for the scaling dimension of another
$SU(12)$-breaking operator.
\end{abstract}
\subjectindex{B31, B37, B87, B38, B44}
\maketitle

\section{Introduction and result}
\label{sec:1}
Four-dimensional conformal field theories that may be realized as a low-energy
limit of a non-Abelian gauge theory with $N$~flavor massless
fermions~\cite{Banks:1981nn} are of great interest phenomenologically because
they can be a starting point for finding viable models of the walking
technicolor~\cite{Holdom:1981rm,Holdom:1984sk,Yamawaki:1985zg,%
Appelquist:1986an,Appelquist:1986tr,Appelquist:1987fc}. Recognition that a
non-perturbative study of such conformal theories is feasible with
currently available lattice techniques~\cite{Appelquist:2007hu} triggered many
recent investigations; see a recent review~\cite{Giedt:2015alr} and the
references cited therein. Here, one is particularly interested in the mass
anomalous dimension of the fermion, $\gamma_m$, which must be of order one in
viable technicolor models. 

It is always challenging, however, to determine something quantitative for a
conformal field theory by lattice numerical simulations. This is natural
because the conformal field theory has no specific length scale and
consequently one ideally has to work with an infinite volume.\footnote{An
intriguing possibility to evade this is to employ the conformal mapping
from~$\mathbb{R}^4$ to~$\mathbb{R}\times\mathbb{S}^3$ and a lattice
discretization of the latter space~\cite{Brower:2016moq}. See also
Ref.~\cite{Ishikawa:2013wf} for an alternative approach.} In fact, for example,
although there seems to be a consensus that the $SU(3)$ gauge theory with
$12$~fundamental massless fermions---$12$-flavor quantum chromodynamics
(QCD)---has an infrared fixed point, there still exist large discrepancies
among central values of the mass anomalous dimension at the fixed point,
$\gamma_m^*$, depending on computational strategies; see Fig.~11
of~Ref.~\cite{Itou:2013kaa} and Table~4 of~Ref.~\cite{Giedt:2015alr}.

Originally motivated by the above large discrepancies in~$\gamma_m^*$, in this
paper we apply the numerical conformal bootstrap---a powerful rigorous
approach to higher-dimensional conformal field theories---to four-dimensional
conformal field theories with an $SU(N)$ global symmetry. A partial list of
references on the numerical conformal bootstrap is~\cite{Rattazzi:2008pe,%
Rychkov:2009ij,Poland:2010wg,Rattazzi:2010gj,Rattazzi:2010yc,Poland:2011ey,%
ElShowk:2012ht,Liendo:2012hy,ElShowk:2012hu,Beem:2013qxa,Kos:2013tga,%
El-Showk:2013nia,Berkooz:2014yda,El-Showk:2014dwa,Nakayama:2014lva,%
Nakayama:2014yia,Alday:2014qfa,Chester:2014fya,Kos:2014bka,Nakayama:2014sba,%
Beem:2014zpa,Chester:2014gqa,Simmons-Duffin:2015qma,Gliozzi:2015qsa}; see also
a most recent paper, Ref.~\cite{Nakayama:2016jhq}, and the recent
review~\cite{Simmons-Duffin:2016gjk} for a more complete list. Our formulation
is valid for arbitrary~$N$, but we will report our numerical results only
for~$N=12$ in the main text (we present the results for~$N=8$ and~$N=16$
in~Appendix~\ref{app:A}). As explained below, by combining a result from our
numerical conformal bootstrap and the fact that lattice simulations of the
$12$-flavor QCD~\cite{Fodor:2011tu,Appelquist:2011dp,DeGrand:2011cu,%
Cheng:2011ic,Aoki:2012eq,Cheng:2013eu,Itou:2013kaa,Cheng:2013xha,%
Lombardo:2014pda,Itou:2014ota} are consistent with the existence of an infrared
fixed point, we obtain an upper bound on the mass anomalous dimension,
\begin{equation}
   \gamma_m^*\leq1.29,\qquad\text{for $N=12$}.
\label{eq:(1.1)}
\end{equation}
Practically, this upper bound is not so strong, not being able to constrain
values obtained by existing lattice simulations.\footnote{There exists a
rigorous bound that follows from the unitarity~\cite{Mack:1975je},
$\gamma_m^*\leq2$.} Nevertheless,
it appears quite interesting that such a strict bound can be made from very
general properties of a unitary conformal field theory, with additional
information provided by lattice simulations.
There even exists a possibility that this bound might become stronger if the
level of approximations that we made in our numerical conformal bootstrap is
increased.

Now, in the context of the technicolor model, one is interested in the
anomalous dimension of the flavor-singlet scalar density,
\begin{equation}
   S=\sum_{k=1}^N\Bar{\psi}^{\Bar{k}}\psi_k,
\label{eq:(1.2)}
\end{equation}
where $k$ ($\Bar{k}$) denotes the index of the fundamental (anti-fundamental)
representation of~$SU(N)$---the flavor group---in a QCD-like theory. This is
because the expectation value of~$S$ provides the technifermion condensate.
Since the combination~$m_0S$ is not renormalized, $m_0S=mS_R$, where $m_0$ is
the bare mass parameter and the right-hand side is the product of the
renormalized quantities, the anomalous dimension of~$S$ is given by the mass
anomalous dimension~$\gamma_m$, defined by
\begin{equation}
   \gamma_m=-\left(\mu\frac{\partial}{\partial\mu}\right)_0\ln Z_m,\qquad
   m=Z_m m_0,
\label{eq:(1.3)}
\end{equation}
where the subscript~$0$ implies that bare quantities are kept fixed. We are
interested in the value of~$\gamma_m$ at the infrared fixed point,
$\gamma_m^*$.

In the above QCD-like theory, we assume that the $SU(N)$ flavor group is
chiral in the sense that we actually have the chiral
symmetry~$SU(N)_L\times SU(N)_R$. Then, applying the flavored chiral rotation
to the scalar density~\eqref{eq:(1.2)}, we have a pseudo-scalar density,
\begin{equation}
   \phi_i^{\Bar{k}}
   =\Bar{\psi}^{\Bar{k}}\gamma_5\psi_i
   -\frac{1}{N}\delta_i^{\Bar{k}}
   \sum_{l=1}^N\Bar{\psi}^{\Bar{l}}\gamma_5\psi_l,
\label{eq:(1.4)}
\end{equation}
which belongs to the adjoint representation of~$SU(N)$. Since the flavor
rotation and the scale transformation commute, the pseudo-scalar adjoint
operator~$\phi_i^{\Bar{k}}$ possesses the same scaling
dimension~$\Delta_{\phi_i^{\Bar{k}}}$ as $S$~\eqref{eq:(1.2)}. Then, the mass
anomalous dimension~$\gamma_m^*$ and the scaling
dimension~$\Delta_{\phi_i^{\Bar{k}}}$ (at the fixed point) are related by
\begin{equation}
   \gamma_m^*=3-\Delta_{\phi_i^{\Bar{k}}}.
\label{eq:(1.5)}
\end{equation}
This also directly follows from the partially conserved axial current (PCAC)
relation.

In Sect.~\ref{sec:2}, we consider a four-point function of a spin~$0$
adjoint operator~$\phi_i^{\Bar{k}}$ without specifying its actual microscopic
structure such as~Eq.~\eqref{eq:(1.4)}.\footnote{We do not assume the
underlying gauge theory either; we assume only that the theory is conformal
and possesses a global~$SU(N)$ symmetry.} We derive the crossing relation
associated with the four-point function,\footnote{We learned that this crossing
relation had already been derived in~Ref.~\cite{Berkooz:2014yda}. We would like
to thank the referee for pointing out this fact.} basically following the
notational conventions of~Ref.~\cite{Poland:2011ey}. Then,
in~Sect.~\ref{sec:3}, we apply the numerical conformal bootstrap to the
crossing relation. For this, we used a semidefinite programming code, the SDPB
of~Ref.~\cite{Simmons-Duffin:2015qma}.

In this way, among other things, we found that for~$N=12$ the system contains a
spin~$0$ relevant operator in the representation~$[N-1,N-1,1,1]$
of~$SU(N)$,\footnote{We label representations of~$SU(N)$ by a list of the
(non-increasing) number of boxes in each column of the corresponding Young
tableau. For example, the adjoint representation is denoted as~$[N-1,1]$.
For~$N=12$, we should say $[11,11,1,1]$ rather than $[N-1,N-1,1,1]$, but in
this paper we use the latter notation even for~$N=12$. This remark applies also
for other representations and for other values of~$N$.} when
\begin{equation}
   \Delta_{\phi_i^{\Bar{k}}}<1.71,\qquad\text{for $N=12$}.
\label{eq:(1.6)}
\end{equation}
Since this relevant operator in the $[N-1,N-1,1,1]$ representation appears in
the operator product expansion (OPE) of two $\phi_i^{\Bar{k}}$s, if the latter
is identified with the pseudo-scalar density in~Eq.~\eqref{eq:(1.4)}, this is a
scalar density. Such an $SU(12)$ non-invariant operator is not radiatively
induced, even if it is relevant, \emph{if\/} our regularization preserves the
$SU(12)$ symmetry. We note, however, that in all existing lattice simulations
of the $12$-flavor QCD, the staggered fermion~\cite{Susskind:1976jm} is
employed to prevent the fermion mass operator (which is believed to be a unique
spin~$0$ $SU(12)$-invariant relevant operator associated with the infrared
fixed point under consideration) from being radiatively induced. This is
accomplished by the exact $U(1)_A$ symmetry~\cite{Kawamoto:1981hw} that the
massless staggered fermion possesses. Still, however, the staggered fermion
cannot preserve the full $SU(12)$ flavor symmetry (the so-called taste
breaking). Generally, when the regularization does not preserve a symmetry,
relevant operators that are not invariant under the symmetry are radiatively
induced and, to achieve the desired continuum or low-energy limit, one has to
tune the coefficients of those non-invariant operators in the action. The fact
that actual lattice simulations~\cite{Fodor:2011tu,Appelquist:2011dp,%
DeGrand:2011cu,Cheng:2011ic,Aoki:2012eq,Cheng:2013eu,Itou:2013kaa,%
Cheng:2013xha,Lombardo:2014pda,Itou:2014ota} of the $12$-flavor QCD are
consistent with the existence of an infrared fixed point without such a
fine-tuning strongly indicates that the theory does not contain the above
$SU(12)$ non-invariant relevant operator in the spectrum.

Thus, assuming the absence of the spin~$0$ relevant operator in the
representation~$[N-1,N-1,1,1]$, we have the
inequality~$\Delta_{\phi_i^{\Bar{k}}}\geq1.71$. Then the upper bound on the mass
anomalous dimension~\eqref{eq:(1.1)} follows from the
relation~\eqref{eq:(1.5)}.

We stress that our upper bound~\eqref{eq:(1.1)} is a physical property of a 
conformal field theory at the infrared fixed point under consideration. The
validity of our upper bound and whether one uses the staggered fermion in
actual lattice simulations are completely independent issues. We have used the
fact indicated by existing lattice simulations, just to support our assumption
on the absence of the spin~$0$ relevant operator in the
representation~$[N-1,N-1,1,1]$ around the fixed point. Whether there exists
such a relevant operator in the RG flow near a fixed point or not is a property
of the fixed point and this property should be independent of the way one
studies the system.

To really claim that the $SU(12)$ non-invariant operator in the $[N-1,N-1,1,1]$
representation is induced with the staggered fermion, we still have to show
that it is not prohibited by exact symmetries of the staggered
fermion~\cite{Golterman:1984cy,Golterman:1984dn}. This group-theoretical
question can be studied with the help of~Ref.~\cite{Lee:1999zxa}, which
provides a complete list of $SU(12)$ non-invariant\footnote{This reference
studies the $SU(4)$ case but we can simply triple the results for $SU(12)$.}
operators up to the canonical mass dimension~$6$; these are consistent with
(i.e., not prohibited by) exact symmetries of the staggered fermion. The
authors of~Ref.~\cite{Lee:1999zxa} show that, for example, the following
four-Fermi scalar operator is consistent with exact symmetries of the
staggered fermion:
\begin{equation}
   X\equiv\sum_{\mu=1}^4
   \sum_{k,i=1}^{12}\Bar{\psi}^{\Bar{k}}\gamma_\mu(\xi_5)_k^{\Bar{i}}\psi_i
   \sum_{l,j=1}^{12}\Bar{\psi}^{\Bar{l}}\gamma_\mu(\xi_5)_l^{\Bar{j}}\psi_j,
\label{eq:(1.7)}
\end{equation}
where $\gamma_\mu$ is the conventional Dirac matrix and $\xi_5$ is a
flavor-space counterpart of the $\gamma_5$ matrix. To examine whether this
combination contains the $[N-1,N-1,1,1]$ representation under the decomposition
into irreducible representations of~$SU(12)$, we take a possible explicit form
of an operator in the $[N-1,N-1,1,1]$ representation,
\begin{equation}
   \mathcal{O}_{(ij)}^{(\Bar{k}\Bar{l})}
   =\left[\Bar{\psi}^{(\Bar{k}}\psi_{(i}
   -\frac{1}{N}\delta_{(i}^{(\Bar{k}}\sum_{m=1}^N\Bar{\psi}^{\Bar{m}}\psi_m\right]
   \left[\Bar{\psi}^{\Bar{l})}\psi_{j)}
   -\frac{1}{N}\delta_{j)}^{\Bar{l})}\sum_{n=1}^N\Bar{\psi}^{\Bar{n}}\psi_n\right],
\label{eq:(1.8)}
\end{equation}
where $(\phantom{ij})$ stands for the symmetrization of the indices enclosed,
and consider the two-point function
\begin{equation}
   \left\langle X\mathcal{O}_{(ij)}^{(\Bar{k}\Bar{l})}\right\rangle
\label{eq:(1.9)}
\end{equation}
in the system of \emph{free fermions}. If this two-point function is non-zero,
then the operator~$X$ contains the component of the $[N-1,N-1,1,1]$
representation. Assuming a particular representation of~$\xi_5$ in which the
component~$(\xi_5)_1^{\Bar{1}}$ is non-zero, it is easy to see that
$\langle X\mathcal{O}_{(11)}^{(\Bar{1}\Bar{1})}\rangle\propto-32(1-2/N+4/N^2)$.
This shows the above assertion: Exact symmetries of the staggered fermion
cannot exclude the relevant operator in the $[N-1,N-1,1,1]$ representation
of~$SU(12)$ from being radiatively induced.

\section{$SU(N)$ crossing relation}
\label{sec:2}
As noted in the previous section, we consider a four-point correlation function
of a spin~$0$ operator in the adjoint representation of the global
symmetry~$SU(N)$,
\begin{equation}
   \left\langle
   \phi_i^{\Bar{k}}(x_1)\phi_j^{\Bar{l}}(x_2)
   \phi_a^{\Bar{c}}(x_3)\phi_b^{\Bar{d}}(x_4)\right\rangle,
\label{eq:(2.1)}
\end{equation}
where the lower (upper) indices stand for indices of the fundamental
(anti-fundamental) representation of~$SU(N)$. In what follows, the scaling
dimension of~$\phi_i^{\Bar{k}}$, $\Delta_{\phi_i^{\Bar{k}}}$, is also denoted
as~$d$:
\begin{equation}
   d\equiv\Delta_{\phi_i^{\Bar{k}}}.
\label{eq:(2.2)}
\end{equation}

In the conformal field theory, four-point functions such
as~Eq.~\eqref{eq:(2.1)} can be computed by applying the OPE to pairs of
operators. The OPE between two operators in the adjoint representation
of~$SU(N)$ is decomposed into the sum over operators in various irreducible
representations of~$SU(N)$ (the Clebsch--Gordon decomposition) as
\begin{align}
   \phi_i^{\Bar{k}}\times \phi_j^{\Bar{l}}
   &\sim
   \sum_{[N-1,N-1,1,1]^+}\mathcal{O}_{(ij)}^{(\Bar{k}\Bar{l})}
   +\sum_{[N-2,1,1]^-}\mathcal{O}_{(ij)}^{[\Bar{k}\Bar{l}]}
   +\sum_{\overline{[N-2,1,1]}^-}\mathcal{O}_{[ij]}^{(\Bar{k}\Bar{l})}
   +\sum_{[N-2,2]^+}\mathcal{O}_{[ij]}^{[\Bar{k}\Bar{l}]}
\notag\\
   &\qquad{}
   +\sum_{[N-1,1]^+}\left[
   \delta_i^{\Bar{l}}\mathcal{O}_j^{\Bar{k}}
   +\delta_j^{\Bar{k}}\mathcal{O}_i^{\Bar{l}}
   -\frac{2}{N}
   \left(
   \delta_i^{\Bar{k}}\mathcal{O}_j^{\Bar{l}}
   +\delta_j^{\Bar{l}}\mathcal{O}_i^{\Bar{k}}
   \right)
   \right]
\notag\\
   &\qquad\qquad{}
   +\sum_{[N-1,1]^-}\left(
   \delta_i^{\Bar{l}}\mathcal{O}_j^{\Bar{k}}
   -\delta_j^{\Bar{k}}\mathcal{O}_i^{\Bar{l}}
   \right)
\notag\\
   &\qquad\qquad\qquad{}
   +\sum_{1^+}\left(
   \delta_i^{\Bar{l}}\delta_j^{\Bar{k}}
   -\frac{1}{N}\delta_i^{\Bar{k}}\delta_j^{\Bar{l}}
   \right)
   \mathcal{O}.
\label{eq:(2.3)}
\end{align}
In this expression, $(\phantom{ij})$ and $[\phantom{ij}]$ stand for the
symmetrization and anti-symmetrization of the indices enclosed and all
operators are traceless with respect to any pair of upper and lower indices. We
label irreducible representations of~$SU(N)$ by a list of the number of boxes
in each column of the corresponding Young tableau. The bar stands for the
conjugate representation and the $1$ in the last term stands for the singlet
representation. The dimensions of each representation are,
$N^2(N-1)(N+3)/4$,
$(N^2-1)(N^2-4)/4$,
$(N^2-1)(N^2-4)/4$,
$N^2(N+1)(N-3)/4$,
$N^2-1$,
$N^2-1$,
and~$1$, respectively, and thus $(N^2-1)^2$ in total, the dimension of the
product representation on the left-hand side. The $\pm$~sign attached to each
representation denotes the parity of the spin of the operators under the sum.
For example, a spin~$1$ operator in the adjoint representation (there must
exist at least one such operator corresponding to the Noether current
of~$SU(N)$) is included in the third line of the above expression ($[N-1,1]^-$).

First we apply the OPE~\eqref{eq:(2.3)} to~Eq.~\eqref{eq:(2.1)} as follows:
\begin{equation}
   \left\langle
   \contraction{}{\phi}{_i^{\Bar{k}}(x_1)}{\phi}
   \phi_i^{\Bar{k}}(x_1)\phi_j^{\Bar{l}}(x_2)
   \contraction{}{\phi}{_a^{\Bar{c}}(x_3)}{\phi}
   \phi_a^{\Bar{c}}(x_3)\phi_b^{\Bar{d}}(x_4)
   \right\rangle.
\label{eq:(2.4)}
\end{equation}
Then, we have
\begin{align}
   &x_{12}^{2d}x_{34}^{2d}\left\langle
   \phi_i^{\Bar{k}}(x_1)\phi_j^{\Bar{l}}(x_2)
   \phi_a^{\Bar{c}}(x_3)\phi_b^{\Bar{d}}(x_4)\right\rangle
\notag\\
   &=
   \sum_{[N-1,N-1,1,1]^+}
   \lambda_{\mathcal{O}}^2
   T{}_{(ij)}^{(\Bar{k}\Bar{l})}{}_{(ab)}^{(\Bar{c}\Bar{d})}
   g_{\Delta,\ell}(u,v)
\notag\\
   &\qquad{}
   +\sum_{[N-2,1,1]^-}
   \lambda_{\mathcal{O}}^2
   \left(
   T{}_{(ij)}^{[\Bar{k}\Bar{l}]}{}_{[ab]}^{(\Bar{c}\Bar{d})}
   +T{}_{[ij]}^{(\Bar{k}\Bar{l})}{}_{(ab)}^{[\Bar{c}\Bar{d}]}
   \right)
   g_{\Delta,\ell}(u,v)
\notag\\
   &\qquad{}
   +\sum_{[N-2,2]^+}
   \lambda_{\mathcal{O}}^2
   T{}_{[ij]}^{[\Bar{k}\Bar{l}]}{}_{[ab]}^{[\Bar{c}\Bar{d}]}
   g_{\Delta,\ell}(u,v)
\notag\\
   &\qquad{}
   +\sum_{[N-1,1]^+}
   \lambda_{\mathcal{O}}^2
   \biggl(
   \delta_i^{\Bar{l}}\delta_a^{\Bar{d}}
   \left(\delta_j^{\Bar{c}}\delta_b^{\Bar{k}}
   -\frac{1}{N}\delta_j^{\Bar{k}}\delta_b^{\Bar{c}}\right)
   +\delta_i^{\Bar{l}}\delta_b^{\Bar{c}}
   \left(\delta_j^{\Bar{d}}\delta_a^{\Bar{k}}
   -\frac{1}{N}\delta_j^{\Bar{k}}\delta_a^{\Bar{d}}\right)
\notag\\
   &\qquad\qquad\qquad\qquad\qquad{}
   -\frac{2}{N}
   \left[
   \delta_i^{\Bar{l}}\delta_a^{\Bar{c}}
   \left(\delta_j^{\Bar{d}}\delta_b^{\Bar{k}}
   -\frac{1}{N}\delta_j^{\Bar{k}}\delta_b^{\Bar{d}}\right)
   +\delta_i^{\Bar{l}}\delta_b^{\Bar{d}}
   \left(\delta_j^{\Bar{c}}\delta_a^{\Bar{k}}
   -\frac{1}{N}\delta_j^{\Bar{k}}\delta_a^{\Bar{c}}\right)
   \right]
\notag\\
   &\qquad\qquad\qquad\qquad\qquad{}
   +\delta_j^{\Bar{k}}\delta_a^{\Bar{d}}
   \left(\delta_i^{\Bar{c}}\delta_b^{\Bar{l}}
   -\frac{1}{N}\delta_i^{\Bar{l}}\delta_b^{\Bar{c}}\right)
   +\delta_j^{\Bar{k}}\delta_b^{\Bar{c}}
   \left(\delta_i^{\Bar{d}}\delta_a^{\Bar{l}}
   -\frac{1}{N}\delta_i^{\Bar{l}}\delta_a^{\Bar{d}}\right)
\notag\\
   &\qquad\qquad\qquad\qquad\qquad\qquad{}
   -\frac{2}{N}
   \left[
   \delta_j^{\Bar{k}}\delta_a^{\Bar{c}}
   \left(\delta_i^{\Bar{d}}\delta_b^{\Bar{l}}
   -\frac{1}{N}\delta_i^{\Bar{l}}\delta_b^{\Bar{d}}\right)
   +\delta_j^{\Bar{k}}\delta_b^{\Bar{d}}
   \left(\delta_i^{\Bar{c}}\delta_a^{\Bar{l}}
   -\frac{1}{N}\delta_i^{\Bar{l}}\delta_a^{\Bar{c}}\right)
   \right]
\notag\\
   &\qquad\qquad\qquad\qquad\qquad{}
   -\frac{2}{N}\biggl\{
   \delta_i^{\Bar{k}}\delta_a^{\Bar{d}}
   \left(\delta_j^{\Bar{c}}\delta_b^{\Bar{l}}
   -\frac{1}{N}\delta_j^{\Bar{l}}\delta_b^{\Bar{c}}\right)
   +\delta_i^{\Bar{k}}\delta_b^{\Bar{c}}
   \left(\delta_j^{\Bar{d}}\delta_a^{\Bar{l}}
   -\frac{1}{N}\delta_j^{\Bar{l}}\delta_a^{\Bar{d}}\right)
\notag\\
   &\qquad\qquad\qquad\qquad\qquad\qquad{}
   -\frac{2}{N}
   \left[
   \delta_i^{\Bar{k}}\delta_a^{\Bar{c}}
   \left(\delta_j^{\Bar{d}}\delta_b^{\Bar{l}}
   -\frac{1}{N}\delta_j^{\Bar{l}}\delta_b^{\Bar{d}}\right)
   +\delta_i^{\Bar{k}}\delta_b^{\Bar{d}}
   \left(\delta_j^{\Bar{c}}\delta_a^{\Bar{l}}
   -\frac{1}{N}\delta_j^{\Bar{l}}\delta_a^{\Bar{c}}\right)
   \right]
   \biggr\}
\notag\\
   &\qquad\qquad\qquad\qquad\qquad{}
   -\frac{2}{N}\biggl\{
   \delta_j^{\Bar{l}}\delta_a^{\Bar{d}}
   \left(\delta_i^{\Bar{c}}\delta_b^{\Bar{k}}
   -\frac{1}{N}\delta_i^{\Bar{k}}\delta_b^{\Bar{c}}\right)
   +\delta_j^{\Bar{l}}\delta_b^{\Bar{c}}
   \left(\delta_i^{\Bar{d}}\delta_a^{\Bar{k}}
   -\frac{1}{N}\delta_i^{\Bar{k}}\delta_a^{\Bar{d}}\right)
\notag\\
   &\qquad\qquad\qquad\qquad\qquad\qquad{}
   -\frac{2}{N}
   \left[
   \delta_j^{\Bar{l}}\delta_a^{\Bar{c}}
   \left(\delta_i^{\Bar{d}}\delta_b^{\Bar{k}}
   -\frac{1}{N}\delta_i^{\Bar{k}}\delta_b^{\Bar{d}}\right)
   +\delta_j^{\Bar{l}}\delta_b^{\Bar{d}}
   \left(\delta_i^{\Bar{c}}\delta_a^{\Bar{k}}
   -\frac{1}{N}\delta_i^{\Bar{k}}\delta_a^{\Bar{c}}\right)
   \right]
   \biggr\}
   \biggr)
\notag\\
   &\qquad\qquad\qquad\qquad\qquad\qquad\qquad\qquad{}
   \times
   g_{\Delta,\ell}(u,v)
\notag\\
   &\qquad{}
   +\sum_{[N-1,1]^-}
   \lambda_{\mathcal{O}}^2
   \biggl[
   \delta_i^{\Bar{l}}\delta_a^{\Bar{d}}
   \left(
   \delta_j^{\Bar{c}}\delta_b^{\Bar{k}}
   -\frac{1}{N}\delta_j^{\Bar{k}}\delta_b^{\Bar{c}}
   \right)
   -\delta_i^{\Bar{l}}\delta_b^{\Bar{c}}
   \left(
   \delta_j^{\Bar{d}}\delta_a^{\Bar{k}}
   -\frac{1}{N}\delta_j^{\Bar{k}}\delta_a^{\Bar{d}}
   \right)
\notag\\
   &\qquad\qquad\qquad\qquad\qquad{}
   -\delta_j^{\Bar{k}}\delta_a^{\Bar{d}}
   \left(
   \delta_i^{\Bar{c}}\delta_b^{\Bar{l}}
   -\frac{1}{N}\delta_i^{\Bar{l}}\delta_b^{\Bar{c}}
   \right)
   +\delta_j^{\Bar{k}}\delta_b^{\Bar{c}}
   \left(
   \delta_i^{\Bar{d}}\delta_a^{\Bar{l}}
   -\frac{1}{N}\delta_i^{\Bar{l}}\delta_a^{\Bar{d}}
   \right)
   \biggr]
   g_{\Delta,\ell}(u,v)
\notag\\
   &\qquad{}
   +\sum_{1^+}
   \lambda_{\mathcal{O}}^2
   \left(
   \delta_i^{\Bar{l}}\delta_j^{\Bar{k}}
   -\frac{1}{N}\delta_i^{\Bar{k}}\delta_j^{\Bar{l}}
   \right)
   \left(
   \delta_a^{\Bar{d}}\delta_b^{\Bar{c}}
   -\frac{1}{N}\delta_a^{\Bar{c}}\delta_b^{\Bar{d}}
   \right)
   g_{\Delta,\ell}(u,v).
\label{eq:(2.5)}
\end{align}
In deriving this, we have used the tensorial structure of the two-point
function of the adjoint operator,
\begin{equation}
   \left\langle
   \mathcal{O}_i^{\Bar{k}}(x)\mathcal{O}_a^{\Bar{c}}(y)\right\rangle
   \propto\left(\delta_i^{\Bar{c}}\delta_a^{\Bar{k}}
   -\frac{1}{N}\delta_i^{\Bar{k}}\delta_a^{\Bar{c}}\right).
\label{eq:(2.6)}
\end{equation}

In Eq.~\eqref{eq:(2.5)}, $\lambda_{\mathcal{O}}$ denotes the OPE coefficient to
a primary operator~$\mathcal{O}$ appearing in the intermediate state;
$\lambda_{\mathcal{O}}$ can be chosen real in unitary conformal field theories.
$\Delta$ and~$\ell$ are the scaling dimension and the spin of the primary
operator~$\mathcal{O}$, respectively. $x_{ij}\equiv x_i-x_j$ and the cross
ratios are defined by
\begin{equation}
   u=\frac{x_{12}^2x_{34}^2}{x_{13}^2x_{24}^2},\qquad
   v=\frac{x_{14}^2x_{23}^2}{x_{13}^2x_{24}^2}.
\label{eq:(2.7)}
\end{equation}
$g_{\Delta,\ell}(u,v)$ is the so-called conformal block and its explicit form in
four dimensions is given by~\cite{Dolan:2003hv}
\begin{align}
   g_{\Delta,\ell}(u,v)
   &=\frac{z\Bar{z}}{z-\Bar{z}}
   \left[k_{\Delta+\ell}(z)k_{\Delta-\ell-2}(\Bar{z})
   -k_{\Delta-\ell-2}(z)k_{\Delta+\ell}(\Bar{z})\right],
\label{eq:(2.8)}
\\
   u&=z\Bar{z},\qquad v=(1-z)(1-\Bar{z}),
\label{eq:(2.9)}
\\
   k_\beta(z)&=z^{\beta/2}{}_2F_1(\beta/2,\beta/2,\beta;z),
\label{eq:(2.10)}
\end{align}
where ${}_2F_1$ is the Gauss hypergeometric function.

Various tensorial symbols appearing in~Eq.~\eqref{eq:(2.5)} are defined by
\begin{align}
   T{}_{(ij)}^{(\Bar{k}\Bar{l})}{}_{(ab)}^{(\Bar{c}\Bar{d})}
   &\equiv
   \delta_{(ij)}^{(\Bar{c}\Bar{d})}\delta_{(ab)}^{(\Bar{k}\Bar{l})}
   -\frac{1}{N+2}
   \left(
   \delta_{(ij)}^{(\Bar{c}\Bar{k})}\delta_{(ab)}^{(\Bar{d}\Bar{l})}
   +\delta_{(ij)}^{(\Bar{c}\Bar{l})}\delta_{(ab)}^{(\Bar{d}\Bar{k})}
   +\delta_{(ij)}^{(\Bar{d}\Bar{k})}\delta_{(ab)}^{(\Bar{c}\Bar{l})}
   +\delta_{(ij)}^{(\Bar{d}\Bar{l})}\delta_{(ab)}^{(\Bar{c}\Bar{k})}
   \right)
\notag\\
   &\qquad{}
   +\frac{2}{(N+1)(N+2)}
   \delta_{(ij)}^{(\Bar{k}\Bar{l})}\delta_{(ab)}^{(\Bar{c}\Bar{d})},
\label{eq:(2.11)}
\\
   T{}_{(ij)}^{[\Bar{k}\Bar{l}]}{}_{[ab]}^{(\Bar{c}\Bar{d})}
   &\equiv
   -\delta_{(ij)}^{(\Bar{c}\Bar{d})}\delta_{[ab]}^{[\Bar{k}\Bar{l}]}
   +\frac{1}{N}
   \left(
   \delta_{(ij)}^{(\Bar{c}\Bar{k})}\delta_{[ab]}^{[\Bar{d}\Bar{l}]}
   -\delta_{(ij)}^{(\Bar{c}\Bar{l})}\delta_{[ab]}^{[\Bar{d}\Bar{k}]}
   +\delta_{(ij)}^{(\Bar{d}\Bar{k})}\delta_{[ab]}^{[\Bar{c}\Bar{l}]}
   -\delta_{(ij)}^{(\Bar{d}\Bar{l})}\delta_{[ab]}^{[\Bar{c}\Bar{k}]}
   \right),
\label{eq:(2.12)}
\\
   T{}_{[ij]}^{(\Bar{k}\Bar{l})}{}_{(ab)}^{[\Bar{c}\Bar{d}]}
   &\equiv
   -\delta_{[ij]}^{[\Bar{c}\Bar{d}]}\delta_{(ab)}^{(\Bar{k}\Bar{l})}
   +\frac{1}{N}
   \left(
   \delta_{[ij]}^{[\Bar{l}\Bar{d}]}\delta_{(ab)}^{(\Bar{k}\Bar{c})}
   -\delta_{[ij]}^{[\Bar{l}\Bar{c}]}\delta_{(ab)}^{(\Bar{k}\Bar{d})}
   +\delta_{[ij]}^{[\Bar{k}\Bar{d}]}\delta_{(ab)}^{(\Bar{l}\Bar{c})}
   -\delta_{[ij]}^{[\Bar{k}\Bar{c}]}\delta_{(ab)}^{(\Bar{l}\Bar{d})}
   \right),
\label{eq:(2.13)}
\\
   T{}_{[ij]}^{[\Bar{k}\Bar{l}]}{}_{[ab]}^{[\Bar{c}\Bar{d}]}
   &\equiv
   \delta_{[ij]}^{[\Bar{c}\Bar{d}]}\delta_{[ab]}^{[\Bar{k}\Bar{l}]}
   -\frac{1}{N-2}
   \left(
   \delta_{[ij]}^{[\Bar{c}\Bar{k}]}\delta_{[ab]}^{[\Bar{d}\Bar{l}]}
   -\delta_{[ij]}^{[\Bar{c}\Bar{l}]}\delta_{[ab]}^{[\Bar{d}\Bar{k}]}
   -\delta_{[ij]}^{[\Bar{d}\Bar{k}]}\delta_{[ab]}^{[\Bar{c}\Bar{l}]}
   +\delta_{[ij]}^{[\Bar{d}\Bar{l}]}\delta_{[ab]}^{[\Bar{c}\Bar{k}]}
   \right)
\notag\\
   &\qquad{}
   +\frac{2}{(N-1)(N-2)}
   \delta_{[ij]}^{[\Bar{k}\Bar{l}]}\delta_{[ab]}^{[\Bar{c}\Bar{d}]},
\label{eq:(2.14)}
\end{align}
and
\begin{equation}
   \delta_{(ij)}^{(\Bar{c}\Bar{d})}\equiv
   \frac{1}{2}
   (\delta_i^{\Bar{c}}\delta_j^{\Bar{d}}+\delta_i^{\Bar{d}}\delta_j^{\Bar{c}}),
\qquad
   \delta_{[ab]}^{[\Bar{k}\Bar{l}]}\equiv
   \delta_a^{\Bar{k}}\delta_b^{\Bar{l}}-\delta_a^{\Bar{l}}\delta_b^{\Bar{k}}.
\label{eq:(2.15)}
\end{equation}
The index structure of these symbols is fixed by the symmetry. The signs are
fixed by requiring positiveness for $i=\Bar{d}$, $j=\Bar{c}$, $\Bar{k}=b$,
and~$\Bar{l}=a$ (see Sect.~2.2 of~Ref.~\cite{Rattazzi:2010yc}, for example).
Noting the identities
\begin{align}
   \delta_{(mj)}^{(\Bar{c}\Bar{m})}&=\frac{1}{2}(N+1)\delta_j^{\Bar{c}},
\label{eq:(2.16)}
\\
   \delta_{[mb]}^{[\Bar{m}\Bar{l}]}&=(N-1)\delta_b^{\Bar{l}},
\label{eq:(2.17)}
\\
   \delta_{(mj)}^{(\Bar{c}\Bar{d})}\delta_{(ab)}^{(\Bar{m}\Bar{l})}
   &=\frac{1}{2}\delta_j^{\Bar{c}}\delta_{(ab)}^{(\Bar{d}\Bar{l})}
   +\frac{1}{2}\delta_j^{\Bar{d}}\delta_{(ab)}^{(\Bar{c}\Bar{l})},
\label{eq:(2.18)}
\\
   \delta_{(mj)}^{(\Bar{c}\Bar{d})}\delta_{[ab]}^{[\Bar{m}\Bar{l}]}
   &=\frac{1}{2}\delta_j^{\Bar{c}}\delta_{[ab]}^{[\Bar{d}\Bar{l}]}
   +\frac{1}{2}\delta_j^{\Bar{d}}\delta_{[ab]}^{[\Bar{c}\Bar{l}]},
\label{eq:(2.19)}
\\
   \delta_{(ij)}^{(\Bar{m}\Bar{d})}\delta_{[mb]}^{[\Bar{k}\Bar{l}]}
   &=-\delta_b^{\Bar{k}}\delta_{(ij)}^{(\Bar{d}\Bar{l})}
   +\delta_b^{\Bar{l}}\delta_{(ij)}^{(\Bar{d}\Bar{k})},
\label{eq:(2.20)}
\end{align}
one can readily confirm that Eq.~\eqref{eq:(2.5)} is consistent with the
tracelessness of the adjoint representation.

Now, in computing the four-point function~\eqref{eq:(2.1)}, we may apply the
OPE~\eqref{eq:(2.3)} in a different order, as
\begin{equation}
   \left\langle
   \contraction{}{\phi}{_i^{\Bar{k}}(x_1)\phi_j^{\Bar{l}}(x_2)
   \phi_a^{\Bar{c}}(x_3)}{\phi}
   \contraction[2ex]{\phi_i^{\Bar{k}}(x_1)}{\phi}{_j^{\Bar{l}}(x_2)}{\phi}
   \phi_i^{\Bar{k}}(x_1)\phi_j^{\Bar{l}}(x_2)
   \phi_a^{\Bar{c}}(x_3)\phi_b^{\Bar{d}}(x_4)
   \right\rangle,
\label{eq:(2.21)}
\end{equation}
which must result in an identical expression. This requirement imposes a strong
consistency condition called the crossing relation. In our case, this is
obtained from the invariance of~Eq.~\eqref{eq:(2.5)} under the exchange
$(x_1,i,\Bar{k})\leftrightarrow(x_3,a,\Bar{c})$. Noting that
$u\leftrightarrow v$ under this exchange, we have, for example, as the
coefficient
of~$\delta_i^{\Bar{k}}\delta_j^{\Bar{l}}\delta_a^{\Bar{c}}\delta_b^{\Bar{d}}$,
\begin{align}
   &\sum_{[N-1,N-1,1,1]^+}
   \lambda_{\mathcal{O}}^2\frac{1}{2(N+1)(N+2)}F_{d,\Delta,\ell}(u,v)
   +\sum_{[N-2,2]^+}
   \lambda_{\mathcal{O}}^2\frac{2}{(N-1)(N-2)}
   F_{d,\Delta,\ell}(u,v)
\notag\\
   &\qquad{}
   +\sum_{[N-1,1]^+}
   \lambda_{\mathcal{O}}^2\frac{-16}{N^3}
   F_{d,\Delta,\ell}(u,v)
   +\sum_{1^+}
   \lambda_{\mathcal{O}}^2\frac{1}{N^2}
   F_{d,\Delta,\ell}(u,v)=0,
\label{eq:(2.22)}
\end{align}
where
\begin{equation}
   F_{d,\Delta,\ell}(u,v)\equiv
   v^dg_{\Delta,\ell}(u,v)-u^dg_{\Delta,\ell}(v,u).
\label{eq:(2.23)}
\end{equation}
We will also use the combination
\begin{equation}
   H_{d,\Delta,\ell}(u,v)\equiv
   v^dg_{\Delta,\ell}(u,v)+u^dg_{\Delta,\ell}(v,u).
\label{eq:(2.24)}
\end{equation}
In a similar way, we have $4!=24$ relations as the coefficients of various
combinations of Kronecker deltas. However, not all the relations are linearly
independent. We find that the linearly independent relations are summarized as
\begin{align}
   &\sum_{[N-1,N-1,1,1]^+}
   \lambda_{\mathcal{O}}^2
   V_{d,\Delta,\ell}^{[N-1,N-1,1,1]^+}
   +\sum_{[N-2,1,1]^-}
   \lambda_{\mathcal{O}}^2V_{d,\Delta,\ell}^{[N-2,1,1]^-}
\notag\\
   &\qquad{}
   +\sum_{[N-2,2]^+}
   \lambda_{\mathcal{O}}^2
   V_{d,\Delta,\ell}^{[N-2,2]^+}
   +\sum_{[N-1,1]^+}
   \lambda_{\mathcal{O}}^2
   V_{d,\Delta,\ell}^{[N-1,1]^+}
\notag\\
   &\qquad\qquad{}
   +\sum_{[N-1,1]^-}
   \lambda_{\mathcal{O}}^2
   V_{d,\Delta,\ell}^{[N-1,1]^-}
   +\sum_{1^+}
   \lambda_{\mathcal{O}}^2
   V_{d,\Delta,\ell}^{1^+}=0,
\label{eq:(2.25)}
\end{align}
where
\begin{align}
   &V_{d,\Delta,\ell}^{[N-1,N-1,1,1]^+}
   \equiv\begin{pmatrix}
   F_{d,\Delta,\ell} \\
   0 \\
   0 \\
   0 \\
   H_{d,\Delta,\ell} \\
   0 \\
   \end{pmatrix},\qquad
   V_{d,\Delta,\ell}^{[N-2,1,1]^-}
   \equiv\begin{pmatrix}
   0 \\
   F_{d,\Delta,\ell} \\
   0 \\
   0 \\
   0 \\
   H_{d,\Delta,\ell} \\
   \end{pmatrix},
\notag\\
   &V_{d,\Delta,\ell}^{[N-2,2]^+}
   \equiv\begin{pmatrix}
   0 \\
   0 \\
   F_{d,\Delta,\ell} \\
   0 \\
   -\frac{4(N-3)(N+1)}{(N-1)(N+3)}H_{d,\Delta,\ell} \\
   \frac{2(N-3)N^2}{(N-2)(N-1)(N+2)}H_{d,\Delta,\ell} \\
   \end{pmatrix},\qquad
   V_{d,\Delta,\ell}^{[N-1,1]^+}
   \equiv\begin{pmatrix}
   0 \\
   0 \\
   0 \\
   F_{d,\Delta,\ell} \\
   -\frac{4(N-2)(N+1)(N+2)}{N^2(N+3)}H_{d,\Delta,\ell} \\
   \frac{N+2}{N}H_{d,\Delta,\ell} \\
   \end{pmatrix},
\notag\\
   &V_{d,\Delta,\ell}^{[N-1,1]^-}
   \equiv\begin{pmatrix}
   -\frac{4(N+1)}{N+2}F_{d,\Delta,\ell} \\
   \frac{2N}{(N-2)(N+2)}F_{d,\Delta,\ell} \\
   \frac{N-1}{N-2}F_{d,\Delta,\ell} \\
   \frac{N^4}{(N-2)^2(N+2)^2}F_{d,\Delta,\ell} \\
    \frac{4(N+1)}{N+3}H_{d,\Delta,\ell} \\
   -\frac{N}{N+2}H_{d,\Delta,\ell} \\
   \end{pmatrix},\qquad
   V_{d,\Delta,\ell}^{1^+}
   \equiv\begin{pmatrix}
   \frac{(N-1)(N+1)}{N(N+2)}F_{d,\Delta,\ell} \\
   \frac{(N-1)(N+1)}{2(N-2)(N+2)}F_{d,\Delta,\ell} \\
   \frac{(N-1)(N+1)}{4(N-2)N}F_{d,\Delta,\ell} \\
   \frac{(N-1)N(N+1)}{(N-2)^2(N+2)^2}F_{d,\Delta,\ell} \\
   -\frac{4(N+1)}{N(N+3)}H_{d,\Delta,\ell} \\
   -\frac{N+1}{N+2}H_{d,\Delta,\ell} \\
   \end{pmatrix}.
\label{eq:(2.26)}
\end{align}

Equation~\eqref{eq:(2.25)} is our crossing relation. It can be confirmed that
the crossing relation~\eqref{eq:(2.25)} we have derived coincides with the
crossing relation in~Ref.~[25] for the same problem [Eqs.~(2.25)--(2.30)
therein], up to the rearrangement of equations and trivial changes in the
notation; this provides a cross-check of our calculation.

The crossing relation~\eqref{eq:(2.25)} restricts possible combinations of the
scaling dimension~$\Delta$, spin~$\ell$, and the OPE
coefficient~$\lambda_{\mathcal{O}}$ of a primary operator~$\mathcal{O}$
appearing in the intermediate state in the four-point function
of~$\phi_i^{\Bar{k}}$, Eq.~\eqref{eq:(2.1)}, whose scaling dimension
is~$d=\Delta_{\phi_i^{\Bar{k}}}$. Besides this constraint, the unitarity requires
$\Delta\geq\Delta_{\text{unitary}}$, where~\cite{Mack:1975je}
\begin{equation}
   \Delta_{\text{unitary}}=\begin{cases}
   1,&\text{for $\ell=0$},\\
   \ell+2,&\text{for $\ell\geq1$},\\
   \end{cases}
\label{eq:(2.27)}
\end{equation}
for a primary operator with the spin~$\ell$ (except the identity operator, for
which~$\Delta=\ell=0$).

\section{Numerical conformal bootstrap}
\label{sec:3}
We now apply the numerical conformal bootstrap to the crossing
relation~\eqref{eq:(2.25)}. We assume that the spin~$0$ adjoint
operator~$\phi_i^{\Bar{k}}$ possesses the smallest scaling
dimension~$d=\Delta_{\phi_i^{\Bar{k}}}$ among all spin~$0$ operators appearing
in~Eq.~\eqref{eq:(2.25)}, except the identity operator for which~$\Delta=0$.

First, we investigate a possible bound on the smallest scaling dimension of a
spin~$0$ operator in the $[N-1,N-1,1,1]$ representation. For this, for a
fixed~$d$, we take an appropriate number~$\Delta_{\text{trial}}\geq d$. Then we
seek a linear differential operator~$\Lambda$, which acts on a $6$-component
vector~$V$ as
\begin{equation}
   \Lambda(V)
   =\sum_{i=1}^6\sum_{1\leq m+n\leq N_{\text{max}}}
   \lambda_{m,n}^i\left.
   \partial_z^m\partial_{\Bar{z}}^n V_i\right|_{z=\Bar{z}=1/2},
\label{eq:(3.1)}
\end{equation}
where coefficients $\lambda_{m,n}^i$ are real, and which fulfills the following
conditions:
\begin{itemize}
\item As a condition for the identity operator for which $\Delta=\ell=0$,
$\Lambda(V_{d,0,0}^{1^+})=1$.
\item As a condition for the spin~$0$ operator in the $[N-1,N-1,1,1]$
representation, $\Lambda(V_{d,\Delta,0}^{[N-1,N-1,1,1]^+})\geq0$
for any~$\Delta\geq\Delta_{\text{trial}}$.
\item For higher-spin~$\ell>0$ operators in the $[N-1,N-1,1,1]$ representation,
$\Lambda(V_{d,\Delta,\ell}^{[N-1,N-1,1,1]^+})\geq0$
for any~$\Delta\geq\Delta_{\text{unitary}}$.
\item For other representations~$R$, for spin~$0$ operators,
$\Lambda(V_{d,\Delta,0}^{R^+})\geq0$ for any~$\Delta\geq d$.
\item For other representations~$R$, for higher-spin~$\ell>0$ operators,
$\Lambda(V_{d,\Delta,\ell}^{R^\pm})\geq0$ for
any~$\Delta\geq\Delta_{\text{unitary}}$.
\end{itemize}

If we can find a $\Lambda$ which fulfills the above conditions, $\Lambda$
acting on the crossing relation~\eqref{eq:(2.25)} yields a contradiction,
$\text{a strictly positive number}=0$. Thus, we can conclude that, if the
system is a unitary conformal field theory, there \emph{must exist\/} a
spin~$0$ operator in the $[N-1,N-1,1,1]$ representation which possesses the
scaling dimension smaller than the assumed~$\Delta_{\text{trial}}$.
Changing~$\Delta_{\text{trial}}$, we can find a restriction on the scaling
dimension of the spin~$0$ operator in the $[N-1,N-1,1,1]$ representation.

The parameter $N_{\text{max}}$ in~Eq.~\eqref{eq:(3.1)} parametrizes the search
space of~$\Lambda$. When $N_{\text{max}}$ is increased, the possible form
of~$\Lambda$ has more varieties and it becomes easier to find the~$\Lambda$
which fulfills the above conditions. As a consequence, the restriction
on the scaling dimension on the operator becomes stronger when $N_{\text{max}}$
is increased. In our present problem, the upper bound on the mass anomalous
dimension becomes lower when $N_{\text{max}}$ is increased.

\begin{figure}[ht]
\begin{center}
\includegraphics[width=10cm,clip]{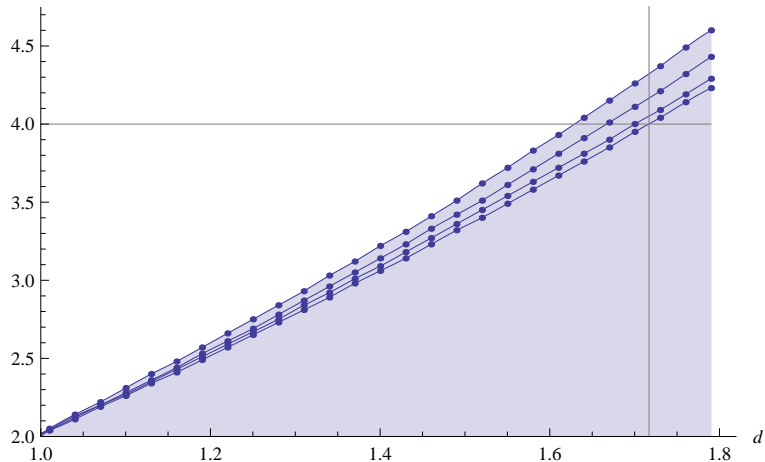}
\caption{Restriction of the smallest scaling dimension of a spin~$0$ operator
in the $[N-1,N-1,1,1]$ representation of~$SU(N)$ with~$N=12$. The horizontal
axis is the scaling dimension of the spin~$0$ adjoint
operator~$\phi_i^{\Bar{k}}$, $d=\Delta_{\phi_i^{\Bar{k}}}$, and the vertical axis
is the scaling dimension of the operator in the $[N-1,N-1,1,1]$ representation.
Boundary curves are obtained by setting, from left to right,
$(\texttt{derivativeOrder}=N_{\text{max}},\texttt{keptPoleOrder},\texttt{Lmax})=%
(10,14,24)$, $(12,14,24)$, $(14,16,24)$, and~$(16,20,24)$, respectively. We see
that the operator becomes relevant, i.e., the scaling dimension becomes smaller
than~$4$, when~$d=\Delta_{\phi_i^{\Bar{k}}}<1.71$.}
\label{fig:1}
\end{center}
\end{figure}

\begin{figure}[ht]
\begin{center}
\includegraphics[width=10cm,clip]{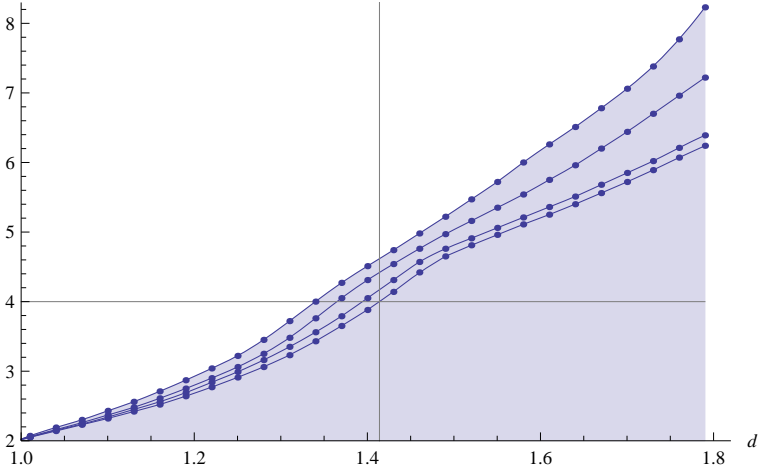}
\caption{Restriction on the smallest scaling dimension of a spin~$0$ operator
in the $[N-2,2]$ representation of~$SU(N)$ with~$N=12$. The horizontal axis is
the scaling dimension of the spin~$0$ adjoint operator~$\phi_i^{\Bar{k}}$,
$d=\Delta_{\phi_i^{\Bar{k}}}$, and the vertical axis is the scaling dimension of
the operator in the $[N-2,2]$ representation. Boundary curves are obtained by
setting, from left to right,
$(\texttt{derivativeOrder}=N_{\text{max}},\texttt{keptPoleOrder},\texttt{Lmax})=%
(10,14,24)$, $(12,14,24)$, $(14,16,24)$, and~$(16,20,26)$, respectively. We see
that the operator becomes relevant when~$d=\Delta_{\phi_i^{\Bar{k}}}<1.41$.}
\label{fig:2}
\end{center}
\end{figure}

The above search for~$\Lambda$ can effectively be carried out by using the
semidefinite programming, as emphasized in~Ref.~\cite{Poland:2011ey}. For this,
we used a semidefinite programming code, SDPB
of~Ref.~\cite{Simmons-Duffin:2015qma}. There are two parameters characterizing
the level of approximation in this approach. One is the maximal spin in the
above search of~$\Lambda$, $\texttt{Lmax}$. Another is the order of the
rational approximation of the conformal block, $\texttt{keptPoleOrder}$. Our
most strict bound below was obtained by setting parameters
as~$(\texttt{derivativeOrder}=N_{\text{max}},\texttt{keptPoleOrder},
\texttt{Lmax})=(16,20,24)$. We confirmed that the boundary curves
in~Figs.~\ref{fig:1} and~\ref{fig:2} do not change, even if we change the
parameters $(\texttt{derivativeOrder},\texttt{keptPoleOrder},\texttt{Lmax})$
to, for example, $(10,11,22)$ for the $N_{\text{max}}=10$ case and to
$(16,18,22)$ (this is only for~Fig.~\ref{fig:1}) and~$(16,18,24)$ for the
$N_{\text{max}}=16$ case.\footnote{For each~$d$, we carry out a binary search to
find the restriction on the scaling dimension of the spin~$0$ operator in
the $[N-1,N-1,1,1]$ representation. We terminate the search when the difference
between two consecutive $\Delta_{\text{trial}}$ becomes less than or equal
to~$0.01$. Thus, we can see the change of the boundary curve only when the
change in the higher is greater than~$0.01$.}

Figure~\ref{fig:1} is our result obtained by the above procedure. The
horizontal axis is the scaling dimension of the spin~$0$ adjoint
operator~$\phi_i^{\Bar{k}}$, $d=\Delta_{\phi_i^{\Bar{k}}}$. The shaded region is
the smallest scaling dimension of a spin~$0$ operator in the $[N-1,N-1,1,1]$
representation of~$SU(N)$ with~$N=12$ in a unitary conformal field theory. We
stress again that to have a unitary conformal field theory, there must exist
at least one spin~$0$ operator in the $[N-1,N-1,1,1]$ representation in the
shaded region. In particular, we see that,
when~$d=\Delta_{\phi_i^{\Bar{k}}}<1.71$, there exists a spin~$0$ relevant (i.e.,
its scaling dimension is smaller than~$4$) operator in the $[N-1,N-1,1,1]$
representation. This leads to our upper bound on the mass anomalous
dimension, Eq.~\eqref{eq:(1.1)}, as explained in~Sect.~\ref{sec:1}.

A similar analysis can be repeated by paying attention to the
representation~$[N-2,2]$ in~Eqs.~\eqref{eq:(2.3)} and~\eqref{eq:(2.25)}.
Figure~\ref{fig:2} is the restriction on the smallest scaling dimension of a
spin~$0$ operator in the $[N-2,2]$ representation of~$SU(N)$ with~$N=12$. This
is obtained by the above numerical conformal bootstrap, by simply exchanging
the role of~$[N-1,N-1,1,1]$ and that of~$[N-2,2]$. We see that there exists a
spin~$0$ relevant operator in the $[N-2,2]$ representation when
$d=\Delta_{\phi_i^{\Bar{k}}}<1.41$. This leads, by repeating the argument
in~Sect.~\ref{sec:1}, to an upper bound on the mass anomalous dimension,
$\gamma_m^*\leq1.59$. This is, however, weaker than the one following from the
$[N-1,N-1,1,1]$ representation, Eq.~\eqref{eq:(1.1)}.

Although our analysis on the representation~$[N-2,2]$ does not provide a useful
upper bound on~$\gamma_m^*$, quite interestingly, we see a kink-like behavior
in the boundary curves in~Fig.~\ref{fig:2}
around~$d=\Delta_{\phi_i^{\Bar{k}}}\sim1.5$. Recalling the fact that in the
numerical conformal bootstrap quite often one finds a known conformal field
theory at a kink point on the boundary curve, the behavior in~Fig.~\ref{fig:2}
is quite suggestive. It would be interesting to study this kink-like behavior
in more detail and seek a possible conformal field theory with a global
$SU(12)$ symmetry that corresponds to the (possible) kink in~Fig.~\ref{fig:2}.

Among other representations in~Eqs.~\eqref{eq:(2.3)}
and~\eqref{eq:(2.25)}, $[N-2,1,1]$ and its conjugate possess only odd spin
operators, and spin~$0$ operators which can correspond to a term in the action
are not included. The representations $[N-1,1]$ and~$1$ are somewhat special
because, depending on the underlying field theory (e.g., $12$-flavor QCD), by
using the flavored chiral rotation it is possible to construct spin~$0$
operators in these representations whose scaling dimension is degenerate
with~$d=\Delta_{\phi_i^{\Bar{k}}}$. For such a case, to draw a non-trivial
conclusion one has to consider the second operator in these representations
that has the scaling dimension greater than or equal to~$d$. Although we carried
out such an analysis for the representations $[N-1,1]$ and~$1$, we do not
present those results here, because the conclusion on the mass anomalous
dimension seems quite dependent on the detail of the underlying theory.

\section*{Acknowledgments}
We are grateful to Tomoki Ohtsuki for an introductory talk on the numerical
conformal bootstrap.
The work of H.~S. is supported in part by Grant-in-Aid for Scientific Research
No.~23540330.

\appendix
\section{Upper bound on~$\gamma_m^*$ for $N=8$ and~$N=16$}
\label{app:A}
Our crossing relation~\eqref{eq:(2.25)} holds for any~$N\geq3$ and, in this
appendix, we present our numerical results for~$N=8$ and~$N=16$. These cases
are also of great interest from perspective of the many-flavor QCD; it is
conceivable that the $SU(3)$ gauge theory with $16$~fundamental massless
fermions is a conformal field theory in the low-energy limit, while whether
$8$-flavor QCD is conformal or not seems not yet quite conclusive; both systems
can be simulated by using the staggered fermion. As for the $N=12$ case in
the main text, we \emph{assume\/} the absence of the spin~$0$ relevant operator
in the representation~$[N-1,N-1,1,1]$ and derive the bound.\footnote{Our result
does not exclude the possibility of the existence of the fixed point
with~$\gamma_m^*>1.33$ [see the bound~\eqref{eq:(A1)}] once we allow the
existence of $SU(8)$-breaking relevant operators. Such a fixed point, if any,
cannot be realized by using the staggered fermion formulation without fine
tuning, but may be realized by the other regularization.}

\begin{figure}[ht]
\begin{center}
\includegraphics[width=10cm,clip]{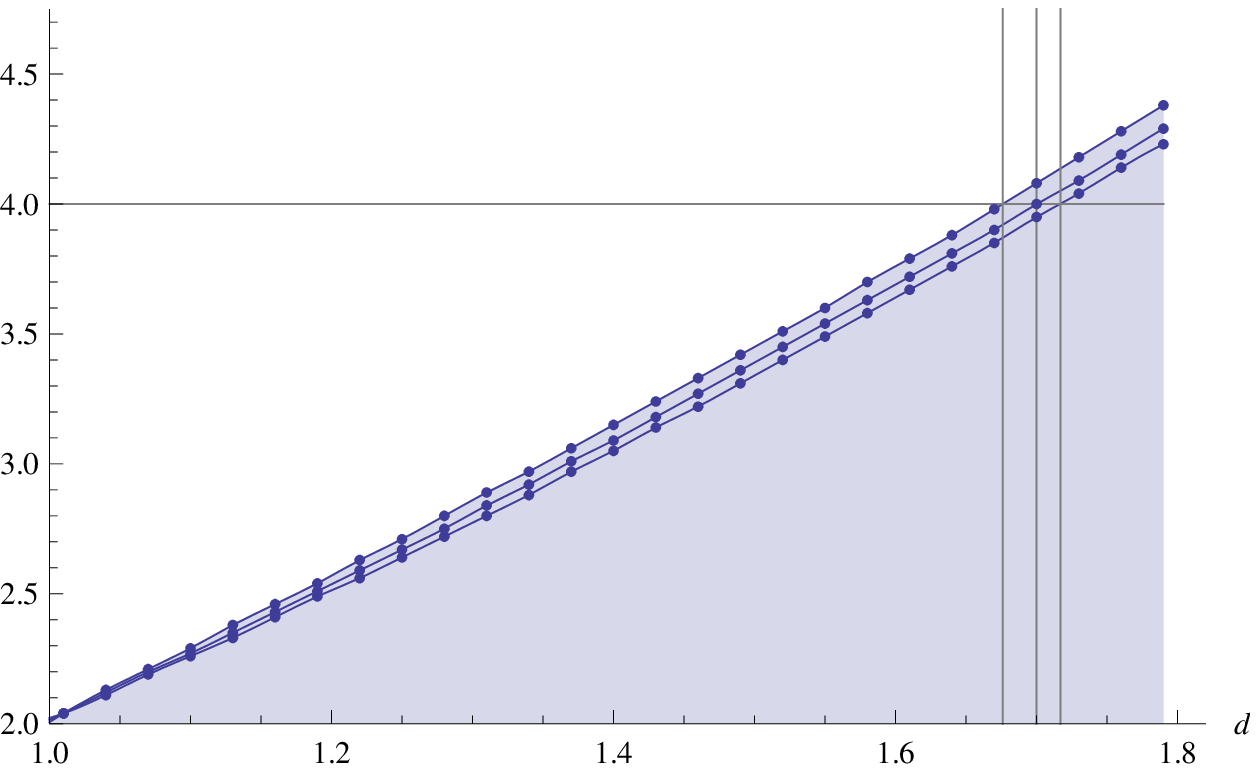}
\caption{Restriction on the smallest scaling dimension of a spin~$0$ operator
in the $[N-1,N-1,1,1]$ representation of~$SU(N)$ with $N=8$, $N=12$,
and~$N=16$ (from left to right). The horizontal axis is the scaling dimension
of the spin~$0$ adjoint operator~$\phi_i^{\Bar{k}}$, $d=\Delta_{\phi_i^{\Bar{k}}}$,
and the vertical axis is the scaling dimension of the operator in the
$[N-1,N-1,1,1]$ representation. We see that the operator becomes relevant
when~$d<1.67$ for~$N=8$, and when~$d<1.71$ for~$N=16$.}
\label{fig:A1}
\end{center}
\end{figure}

Figure~\ref{fig:A1} is our result on the smallest scaling dimension of a
spin~$0$ operator in the $[N-1,N-1,1,1]$ representation of~$SU(N)$ with $N=8$,
$N=12$, and~$N=16$ (from left to right). Boundary curves are obtained by
setting
$(\texttt{derivativeOrder}=N_{\text{max}},\texttt{keptPoleOrder},\texttt{Lmax})=%
(14,16,24)$. As for $N=12$ in the main text, we see that when~$d<1.67$
for~$N=8$, and when~$d<1.71$ for~$N=16$, there emerges an $SU(N)$-breaking
relevant operator in the system. Thus, by assuming the absence of such an
operator, we have an upper bound on the mass anomalous dimension as
\begin{equation}
   \gamma_m^*\leq1.33\qquad\text{for $N=8$},
\label{eq:(A1)}
\end{equation}
and
\begin{equation}
   \gamma_m^*\leq1.29\qquad\text{for $N=16$}.
\label{eq:(A2_)}
\end{equation}
Although the latter bound is numerically the same as~Eq.~\eqref{eq:(1.1)},
which is for~$N=12$, there is no contradiction because here we are using a
somewhat narrower search space for the linear
operator~$\Lambda$ ($N_{\text{max}}=14$) than that in the main text
($N_{\text{max}}=16$); the bound on~$\gamma_m^*$ here is thus somewhat weaker
than would be obtained from the setting in the main text.

\begin{figure}[ht]
\begin{center}
\includegraphics[width=10cm,clip]{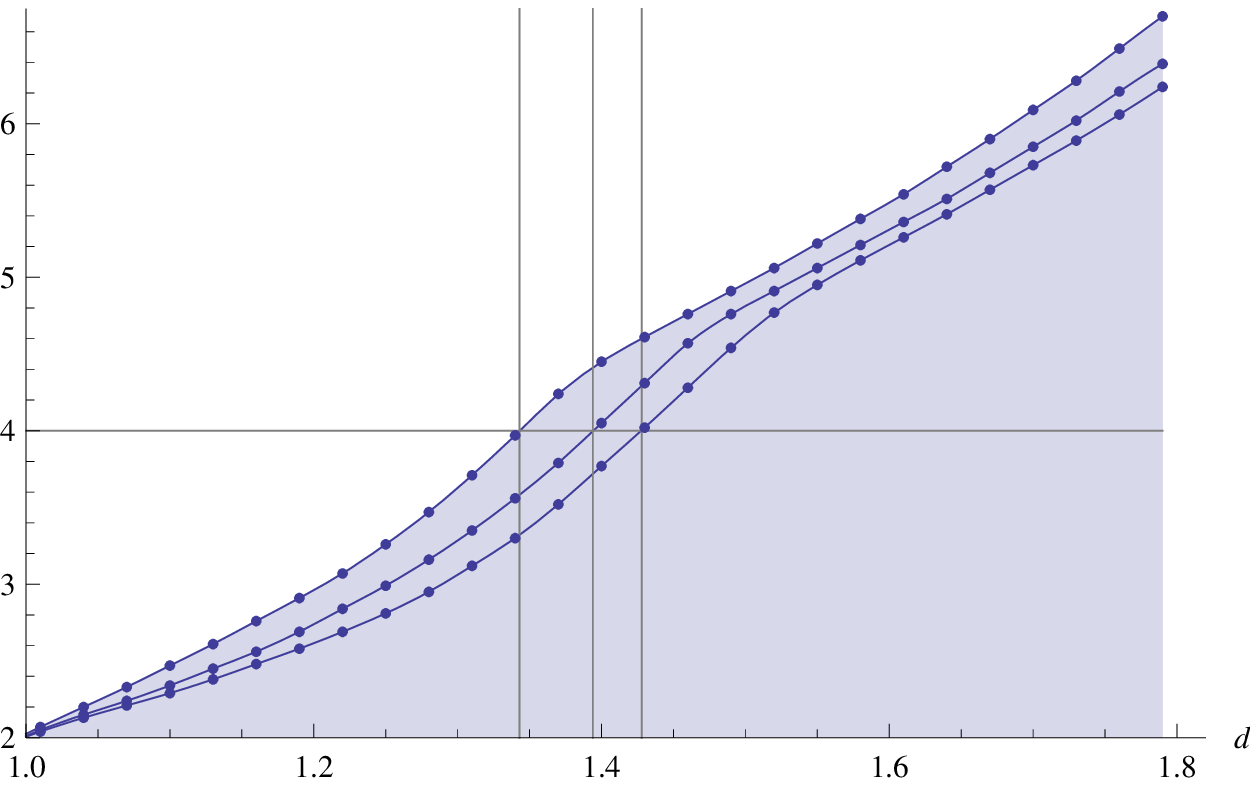}
\caption{Restriction on the smallest scaling dimension of a spin~$0$ operator
in the $[N-2,2]$ representation of~$SU(N)$ with $N=8$, $N=12$, and~$N=16$
(from left to right). The horizontal axis is the scaling dimension of the
spin~$0$ adjoint operator~$\phi_i^{\Bar{k}}$, $d=\Delta_{\phi_i^{\Bar{k}}}$, and
the vertical axis is the scaling dimension of the operator in the $[N-2,2]$
representation. We see that the operator becomes relevant when~$d<1.34$
for~$N=8$, and when~$d<1.42$ for~$N=16$.}
\label{fig:A2}
\end{center}
\end{figure}

Figure~\ref{fig:A2} is our result on the smallest scaling dimension of a
spin~$0$ operator in the $[N-2,2]$ representation of~$SU(N)$ with $N=8$,
$N=12$, and~$N=16$ (from left to right). The parameters
$(\texttt{derivativeOrder}=N_{\text{max}},\texttt{keptPoleOrder},\texttt{Lmax})$
are the same as above. As for the $N=12$ case in the main text, although the
consideration of the operator in the $[N-2,2]$ representation does not provide
a useful bound on~$\gamma_m^*$, we also observe a kink-like behavior for~$N=8$
and~$N=16$. Again, it would be interesting to study this kink-like behavior in
more detail and seek a possible conformal field theory that corresponds to
these (possible) kinks.

\end{document}